# TeV/m Nano-Accelerator: Current Status of CNT-Channeling Acceleration Experiment

Y. M. Shin[1, 2, a)] A. H. Lumpkin[2], J. C. Thangaraj[2], R. M. Thurman-Keup[2], and V. Shiltsev[2]

[1]*Northern Illinois Center for Accelerator and Detector Development (NICADD), Department of Physics, Northern Illinois University*
[2] *Fermi National Accelerator Laboratory, IL 60510.*

[a)]Corresponding author: yshin@niu.edu

**Abstract.** Crystal channeling technology has offered various opportunities in the accelerator community with a viability of ultrahigh gradient (TV/m) acceleration for future HEP collider. The major challenge of channeling acceleration is that ultimate acceleration gradients might require a high power driver in the hard x-ray regime (~ 40 keV). This x-ray energy exceeds those for x-rays as of today, although x-ray lasers can efficiently excite solid plasma and accelerate particles inside a crystal channel. Moreover, only disposable crystal accelerators are possible at such high externally excited fields which would exceed the ionization thresholds destroying the atomic structure, so acceleration will take place only in a short time before full dissociation of the lattice. Carbon-based nanostructures have great potential with a wide range of flexibility and superior physical strength, which can be applied to channeling acceleration. This paper presents a beam-driven channeling acceleration concept with CNTs and discusses feasible experiments with the Advanced Superconducting Test Accelerator (ASTA) in Fermilab.

## INTRODUCTION

The recent discovery of the Higgs boson particle sheds light on the endeavor to explore new particles and new physics beyond the Standard Model which is expected to be unveiled by the LHC at CERN in the near future. A great interest of the scientific community has thus focused on the idea of building a high energy machine, such as ILC, CLIC, LEP3, and Muon Collider based on the belief that a new multi-TeV lepton collider with sufficient luminosity will enable us to understand the mechanism behind mass generation and electroweak symmetry breaking (and possibly to discover supersymmetric particles), and to search for signs of extra space-time dimensions and quantum gravity. The most viable options currently under consideration are the e+e- linear colliders ILC and CLIC. The biggest challenge for the linear colliders is to accelerate particles to the design energy within a reasonable facility footprint and with the maximum possible power conversion from the "wall-plug" to the beams. It is expected that the designed accelerator facilities will cost multiple billions of dollars for construction, operation, and maintenance, even aside from the question of whether the machines are actually capable of providing a sufficient scale of energy resolution. One can safely assume that within an allowable budget scale a future HEP facility should not exceed a few dozen km in length and simultaneously require less than a few dozen MW of beam power. Will such colliders be capable of reaching energies that may be orders of magnitude larger than current ones, namely, 100 – 1000 TeV? Obviously, to obtain the energies of interest within the given footprint, development of ultra-fast acceleration with six-dimensional phase-space beam control will be the most important mission for the future accelerator facilities. The cost models of the modern colliders are quite complicated, but one may safely assume that a future facility should not exceed a few tens of km in length and simultaneously require less than 10 to a few tens of MW of beam. To get to the energies of interest within the given footprint, fast particle acceleration is inevitable.

# HIGH GRADIENT ACCELERATION IN HIGHLY DENSE PLASMA MEDIA

Plasma-wakefield acceleration (PWA) has become of great interest because of the promise to offer extremely large acceleration gradients, on the order of $E_0 \approx n_0^{1/2}$ [GeV/m], where $n_0$ is the ambient electron number density ($n_0$ [$10^{18}$cm$^{-3}$]), on the order of 30-100 GV/m at plasma densities of $n_0 = 10^{17} - 10^{18}$ cm$^{-3}$. The density of charge carriers (conduction electrons) in solids $n_0 = \sim 10^{20} - 10^{23}$ cm$^{-3}$ is significantly higher than what was considered above in plasma, and correspondingly, wakefields of up to 100 GeV/cm or 10 TV/m are possible.

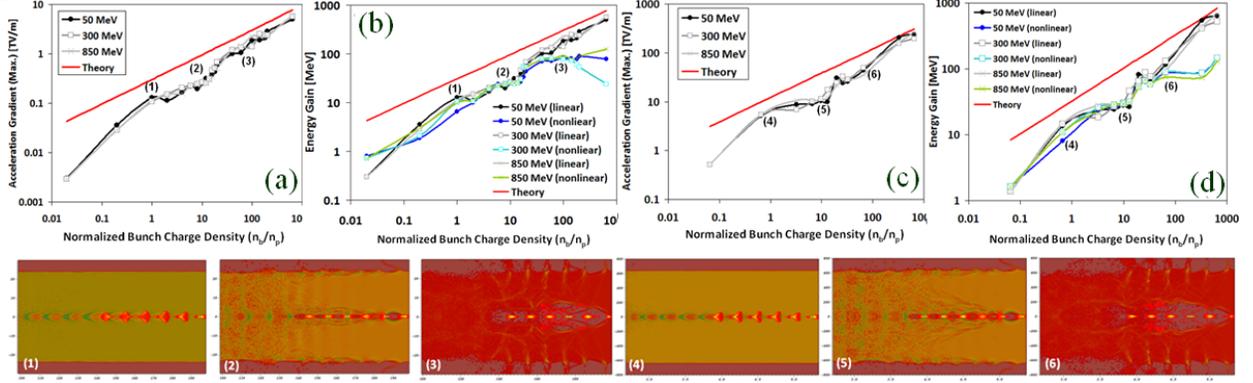

**FIGURE 1.** Top: Acceleration gradients ((a) and (c)) and energy gains ((b) and (d)) versus normalized charge density graphs of multi-bunched beam with 50 MeV, 300 MeV, and 850 MeV with (a)/(b) $n_p = 10^{25}$ m$^{-3}$ and (c)/(d) $n_p = 1.6 \times 10^{28}$ m$^{-3}$, bottom: (1) – (6) corresponding plasma charge distributions.

Figure 1 shows acceleration gradient and energy gain versus beam charge density, normalized by two plasma number densities, $10^{25}$ m$^{-3}$ and $1.6 \times 10^{28}$ m$^{-3}$, which might be in the lower and upper limits of electron density of solid-state media. The corresponding plasma wavelengths are 10 μm and 0.264 μm, which are selected as they are also in the spectral range of available photon driving sources such as IR/UV lasers or magnetic undulators[1]. The bunch charge density, $n_b$, is swept from 0.01 $n_p$ to 1000 $n_p$, ranging from the under-coupled regime to blowout one. With the simulation conditions, 10 micro-bunches are specified to move through a 10$\lambda_p$ thick plasma column. In the graphs, (1) and (4), (2) and (5), and (3) and (6) are at in-between quasi-linear and linear regimes, middle of the linear regime, and in-between linear and blowout regimes with the two density conditions, respectively. The acceleration gradients of a dense plasma state are 4 – 5 orders of magnitude higher than those of plasmas in gas-state. Figures 1(2) and (5) show that the electron bunches in the tails strongly resonate with traveling disturbances in the plasmas, generated by a drive bunch, in agreement with the theoretical prediction. In the solid plasma, as escaping from a driving field due to fast pitch-angle diffusion resulting from increased scattering rates, particles must be accelerated along major crystallographic directions. This is called "channeling acceleration". Normally, crystal channeling has been applied to high energy beam control such as collimation, bending, and refraction[2].

## CNT-CHANNELING ACCELERATION

Wakefields in crystals can be excited by two sorts of driving sources: x-ray laser or short electron bunch. With the x-ray pumping method[3], a crystal channel can hold > $10^{13}$ V/cm transverse and $10^9$ V/cm longitudinal fields of diffracted traveling EM-waves at the Bragg diffraction condition ($\lambda/2b = \sin\theta_B$), where $b$ is the lattice constant and $\theta_B$ is the diffraction angle) However, to hold the ultimate gradients, the acceleration requires coherent hard x-rays ($\hbar\omega \approx 40$ keV) of $\geq 3$ GW to compensate for radiation losses, which exceed those of today. The x-ray driving method thus fits for heavy particles, e.g. muons and protons, which have relatively smaller radiation losses. For electrons, the beam-driven acceleration is more favorably applicable to channeling acceleration as the energy losses of a drive beam can be transformed into acceleration energy of a witness beam[4]. The highly intensive plasma interaction in a crystal channel induces thermal radiations and collisional impacts accompanied by a large amount of heat energy, which would exceed the ionization thresholds and may even destroy the atomic structure. Only disposable forms of crystals such as fibers or films can be used for channeling acceleration. Also, lattice structures of crystals have fixed atomic dimensions, which thereby have some limits in designing acceleration parameters to mitigate physical constraints in solid plasma channels. Carbon nanostructures have potential advantages over crystals for channeling acceleration such as wider channels (weaker de-channeling), broader beams (using nanotube ropes), wider acceptance angles (< 0.1 rad), 3D beam control over greater lengths, and in particular excellent thermal and mechanical strength, which are ideally fit to channeling acceleration and cooling applications, plus beam extraction, steering, and collimation. CNTs are comprised entirely of $sp$2 bonds, which are extremely stable and thermally and mechanically stronger than crystals, steels, and even diamonds ($sp$3 bond). Nanotubes thus have a higher probability of surviving in an extremely intense channeling, radiation, and acceleration environment.

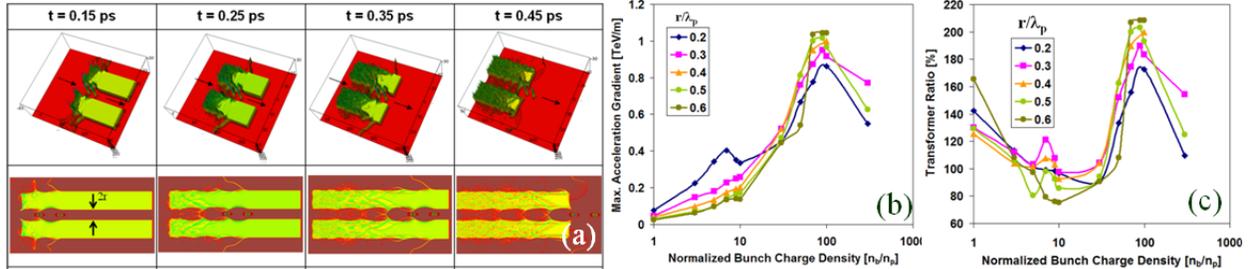

**FIGURE 2.** (a) Time-tagged charge distribution of a hollow plasma acceleration ($n_p = n_b$) with a drive and witness beam (top) 3D charge distribution (middle) 2D distribution (bottom) spatial energy distribution (b) maximum acceleration gradient and (c) transformer ratio versus bunch charge distribution normalized by bunch charge density with various tunnel radii (r = 0.2 – 0.6$\lambda_p$).

Figure 2 shows time-tagged snapshots of a two-beam accelerating system with a ~ 10$\lambda_p$ hollow plasma channel that is modeled with $n_p \sim 10^{25}$ m$^{-3}$. The simulation condition also includes the drive-witness coupling distance of ~ 1.6$\lambda_p$, σ = ~ 0.1$\lambda_p$. and a linear regime bunch charge density of $n_b \sim n_p$. For this simulation, the plasma channel is designed with a tunnel of r = 0.1$\lambda_p$. Just like a uniformly filled one, the drive bunch generates tailing wakes in the hollow channel due to the repulsive space charge force. The plasma waves travel along the hollow channel with the density modulation in the same velocity with the witness beam. Note that the bunch shape of the drive beam in the tunnel remains relatively longer than in the cylindrical plasma column. The energy versus distance plots in Fig. 2 (bottom) shows that sinusoidal energy modulation apparently occurs in the plasma channel perturbed by two bunches. Here, the relative position of the drive beam corresponds to the first maximum energy loss, while that of the witness beam does to the first maximum energy gain. The traveling wakes around the tunnel continuously transform acceleration energy from the drive beam to the witness one. The energy gain and acceleration gradient are fairly limited by the radius and length of the tunnel with respect to plasma wavelength and bunch charge density. Bunch parameters of the beam-driven acceleration system have thus been analyzed with various tunnel radii, as shown in Figs. 2(b) and (c). For the analysis, the bunch charge density was swept from 1 to 300, normalized by plasma density, $n_p$, for five different tunnel radii from 0.2 to 0.6$\lambda_p$ and relativistic beam energy 20 MeV. While in the linear regime $n_b = ~ 1 – 10n_p$, the maximum acceleration gradient drops off with an increase of the tunnel radius from 0.2 to 0.6, it increases in the blowout regime, $n_b = ~ 10 – 100n_p$. The maximum acceleration gradient is increased from ~ 0.82 TeV/m of r = 0.2 $\lambda_p$ to ~ 1.02 TeV/m of r = 0.6 $\lambda_p$ with $n_b = 100n_p$, corresponding to ~ 20 % improvement. The energy transformer ratio follows a similar tendency with the acceleration gradient curve in the linear and blowout regimes. In the linear regime ($n_b/n_p \sim 1 – 10$), scattering is negligibly small, which does not perturb particle distribution of the bunch within the hollow channel. The repulsive space charge force between the bunch and the plasma is increased in the inversely proportional to their spacing. The channeled bunch thus undergoes the higher acceleration gradient as the channel gets smaller. However, in the blowout regime ($n_b/n_p \sim 10 – 100$) the repulsive space charge force from an excessive amount of the bunch charge density against the plasma channel is strong as to heavily perturb the bunch and to scatter electrons out of the bunch. The strength of space charge force is decreased with the channel radius, so the electrons in the bunch is less scattered with an increase of the channel radius. The gradient is thus lowered with an increase of the channel radius accordingly.

## CURRENT EXPERIMENTAL STATUS AT FERMILAB-ASTA

The AAO-CNT (anodic aluminum oxide carbon nanotube) template process is well known nanofabrication technique to implant straight, vertical CNTs in a flat substrate by the chemical vapor deposition (CVD) growth process over a porous aluminum oxide template (Fig. 3(a))[5]. We have been discussing CNT and nanostructure fabrication with the Nanolab Inc in various perspectives. The length of AAO-CNTs is determined by the nano-channel length of the AAO template, i.e. the thickness of AAO films, which can be controlled by tuning the anodizing time. ~ 100 μm thick AAO templates with 20 – 200 nm are commercially available. The outer and inner diameters of AAO CNTs can be tuned by pore sizes on an AAO template and a CVD growth condition, respectively. A 100 μm long, 200 nm wide CNT test sample was first designed and fabricated (Fig. 3(b)).

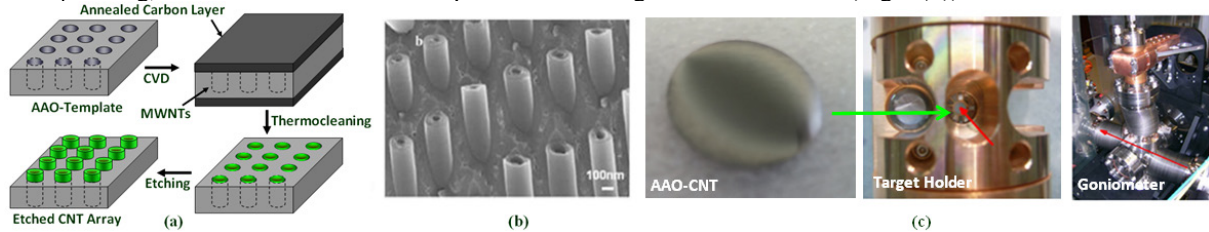

**FIGURE 3.** (a) AAO-CNT process diagram (b) SEM image [ref: 5] (c) AAO-CNT to be in goniometer

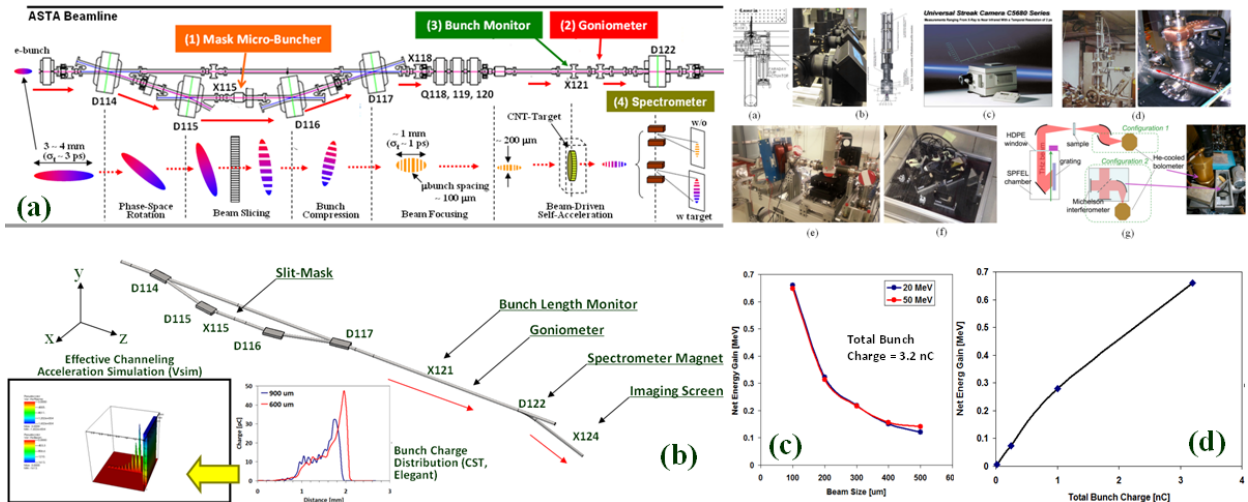

**FIGURE 4.** (a) Details of the beamline in the chicane and after the chicane. X121 station is the designated bunch length monitor and the MPI is located on a table at beam right. The streak camera is located outside the tunnel. Photo images of instruments available at Fermilab (b) ASTA 50 MeV beamline model (CST-PS). Inset is a modulated bunch charge distribution at the goniometer position and effective channeling acceleration model (Vsim[6]). Energy gain versus (c) beam size and (d) bunch charge graphs.

After commissioning the detectors with broadband THz radiation, the CNT target will be installed in the goniometer to assess changes of the beam energy and radiation spectrum due to presence of the target. It should also be noted that the optical transport to the streak camera outside the tunnel is being done with 4 inch diameter mirror optics. Thus THz radiation could also be transported from the Martin-Puplett interferometer (MPI) table to the streak camera optics table outside the tunnel. A cryo-cooled bolometer or other THz detectors with a grating would be used there. At the ASTA 50 MeV beamline, while a ~ 3 ps long photo-electron bunch passes through the bunch compressor (BC1), the slit-mask placed in the BC1 will slice the bunches into micro-bunch trains by imprinting the shadow of a periodic mask onto a bunch with a correlated energy spread. In principle, modulation strength and periodicity of the micro-bunched beam can be controlled by adjusting the grid period or by the dipole magnetic field[7]. According to our theoretical calculations and simulation (Elegant and CST-PS), 900 μm spaced, 300 μm wide slits produce a ~ 100 μm longitudinal beam modulation after BC1. A tungsten slit-mask is already installed in the chicane (BC1, X115), which is ready for a RF-phase scanning test. The experiments will be done at the station X121 located after the chicane as shown in Fig. 4. The autocorrelation data of coherent transition radiation (CTR) employing the MPI will be processed for longitudinal structure and the X121 beam profile monitor should also reveal longitudinal modulation in one transverse plane by powering a skew-quadrupole in the dispersive location in the chicane. In the process of beam-driven self-acceleration, a bunch head (driver) loses energy, while a bunch tail (witness) gains energy. In the transport, an electron bunch is compressed down from $\sigma_t$ ~ 3 ps to ~ 1 ps by the bunch compressor (BC1) and focused by the quadrupole triplet until the beam spot size reaches $\sigma_r$ ~ 100 μm at the goniometer position (X121). The beam is then injected into a target mounted in the goniometer and beam energy is measured by the magnetic spectrometer (D122), as shown in Fig. 3. With the plan to detect wakefield excitation and particle acceleration due to presence of a CNT channel the experiment will be executed by subtracting the beam energy distribution measured with and without a target. Any correlated change with structure angle of the bunch energy distribution projected to the spectrometer screen (X124) before and after loading a target will be a good indicator of wakefield excitation and channeling interaction. More systematic tests are planned to accurately measure the acceleration parameters (energy gain/loss, transformer ratio, etc) of the modulated bunches: first the beam energy will be measured without a target, which will be used as a reference. With a CNT target installed, electron energies will be measured in terms of the beam injection angles with respect to the target axis, as energy gain varies with the channeling angle (= injection angle). Subtracting minimum from maximum projected energy distribution profiles, calibrated by the measured data without a target, will provide the net energy gain of a CNT target. We will sweep the angles with measuring the beam energies by the spectrometer and a difference between maximum and minimum values of the scanned beam energies will equal the net energy gain. With scanning bunch charges and RF-phases, the test will be repeated with a slit-mask at X115 to examine beam-modulation effects on the channeling acceleration. Channeling acceleration under the experimental conditions with real bunch profile (ASTA 50 MeV beamline) and designed CNT-crystalline parameters has been examined by the combined simulation tools employing CST, Elegant, and Vsim (Fig. 4(b)). The full scale beamline from the magnetic chicane to the beam dump is simulated and spatial charge distribution of a 1 ps bunch, modulated by a slit-mask, is monitored at the goniometer position after an image station at X121. The beam profile is manually programmed into the Vsim Beam-Driven Plasma Simulation module designed with an effective CNT channel. The head of the modulated beam quickly excites wakefields along the effective CNT channel, modeled with the equivalent plasma column, and the tail of the beam gains energy from the excited wakefields transforming the driving energy. From parameter sweeps, simulation results predicted 3.2 nC charge with 100-μm beam size produces 0.65 MeV net energy gain over the 100 μm long channel, corresponding to about 6.5 GeV/m (Figs. 4(c) and (d)). The small acceleration gradient of the under-coupled plasma in the quasi-linear regime is mainly attributed to the low charge density of the beam relative to the channel plasma density. Nevertheless, the variation of beam energy due to the presence of the target, if detected in experiment, will assure wakefield excitation in a crystal structure, indicating feasibility of channeling acceleration.

## CONCLUSION AND FUTURE WORK

Despite the great potential of HEP colliders, excessively high driving energy and power requirements accompanied by the insufficient durability of crystal structures has removed channeling acceleration from primary consideration for high gradient (HG) accelerators. However, replacing crystals with nano-structures makes this possible by mitigating the power and energy requirements, with the advantage of improved physical strength. The fundamental mechanisms of plasmon excitation and photon-particle coupling of the nanotube will thus be investigated with Fermilab ASTA electron beamline. Successful demonstration of all the simulations and experimental tests will open new opportunities for HG accelerator research efforts by merging nanotechnology and high energy physics. All of the new techniques and methods developed from heuristics will indeed be incorporated into existing technologies for current HEP collider R&D programs.

## ACKNOWLEDGEMENTS

The authors of this paper would like to thank X. Zhu, D. Broemmelsiek, D. Crawford, D. Mihalcea, D. Still, K. Carlson, J. Santucci, J. Ruan, E. Harms, and P. R. G. Piot for helping this research with simulations and technical support at Fermilab ASTA. This work was supported by the DOE contract No. DEAC02-07CH11359 to the Fermi Research Alliance LLC.